\documentclass[twocolumn,secnumarabic,amssymb, amsmath, nofootinbib,tightenlines,
nobibnotes, aps,
prl]{revtex4-1}
\usepackage{txfonts}
\makeatletter

\newcommand{\Rmnum}[1]{\expandafter\@slowromancap\romannumeral #1@}
\makeatother

\begin{document}


\title{Semi-quantum key distribution protocol using Bell state }


\author{Zhiwei Sun$^1$}
\email[]{sunzhiwei1986@gmail.com}
\author{Ruigang Du$^1$}
\email{Duruigang@yahoo.com.cn}
\author{Dongyang Long$^1$}
\email{issldy@mail.sysu.edu.cn}

\affiliation{
$^{1}$Department of Computer Science, Sun Yat-sen University, Guangzhou 510006,China \\
}


\date{\today}

\begin{abstract}
A quantum key distribution protocol with classical Bob based on polarization entangled photon pairs is presented. It approximates a single photon and exploited the inherent randomness of quantum measurements to attain highly secure keys and high efficiency of the transmission.
\end{abstract}


\maketitle

\section{\Rmnum{1}. INTRODUCTION}
Cryptography is a way to transform information so that it is unintelligible and therefore unless to those who are not meant to have access to it. Thus far, it is trusted that the only proven unconditionally secure crypto-system is the one-time-pad scheme. To employ this scheme, therefore, the two distant communicating parties must have a secure method to share a key that is as long as the message to be encrypted. However, it is not an easy task to share the secret keys between the two parties prior to the communication because they cannnot send a secret key by an open channel to the public. Fortunately Bennett-Brassard (BB84) \cite{BB84} showed how to exploit the properties of quantum mechanics for cryptographic purposes, independently rediscovered by Ekert (E91) \cite{1991Ekert} a few years later, which was the beginning of quantum key distribution (QKD) and have been theoretically proven secure \cite{2000Shor} when both parties are quantum,  Up to now, many quantum key distribution (QKD) protocols have already been proposed \cite{1992确定分发,1992Bennett,1995确定分发,1997确定分发}.

What is possible when only one party (Alice) is quantum, yet the other (Bob) has only classical capabilities? Recently, several "semi-quantum" key distribution protocols (SQKD) were proposed \cite{2007Boyer,2009Boyer,2009Zou}. Boyer et al. \cite{2007Boyer} suggested the idea of semiquantum key distribution using four quantum states. For convenience, we call such a protocol BKM2007. Zou et al. \cite{2009Zou} derived its simplification that requires only one quantum state, ZQLWL2009 for short. These protocols give an answer to how much "quantum" a protocol needs to be in order to achieve a significant advantage over all classical protocols.

The conventional setting when both parties are quantum is as follow: Alice and Bob have labs that are perfectly secure, both of them can perform any quantum operations, and they use qubits for their quantum communication on a quantum channel and also have a classical authenticated channel which can be heard, but cannot be tampered with by the adversary. For SQKD protocol \cite{2007Boyer}, a quantum channel travels from Alice's lab to the outside world and back to her lab. Bob can access a segment of the channel, and whenever a qubit passes through that segment Bob can either let it go undisturbed or $(1)$ measure the qubit in the computational basis $\{|1\rangle, |0\rangle\}$ which is also called "classical" basis; $(2)$ prepare a (fresh) qubit in the classical basis, and send it. Bob is called classical Bob if he is limited to performing only operations $(1)$ and $(2)$ or doing nothing and could never obtain any quantum superposition of the computational basis states. If all the parties are classical, they would always be working with qubits in the classical basis which would then make the resulting protocol be equivalent to an old-fashion classical protocol, and therefore, the operations themselves shall here be considered classical. So this kind of protocol is termed "QKD with classical Bob" or "Semi-quantum key distribution".

The SQKD protocols \cite{2007Boyer,2009Boyer,2009Zou} have been proved being completely robust which is an important step in studying security. Robustness of a protocol means that any attempt of an eavesdropper to obtain information on the INFO string (the definition of INFO string defined in Ref. \cite{2007Boyer}: before Alice and Bob perform the ECC step) necessarily induces some error which is detectable by the legitimate users. In particular, Boyer et al. \cite{2007Boyer} divided robustness into three classes: completely robust, partly robust, and completely nonrobust. A protocol is said to be completely robust if nonzero information acquired by Eve on the INFO string implies nonzero probability that the legitimate participants find errors on the bits tested by the protocol. A protocol is said to be completely nonrobust if Eve can gain the INFO string without inducing any error on the bits tested by the protocol. A protocol is said to be partly robust if Eve can acquire some limited information on the INFO string without inducing any error on the bits tested by the protocol. Partly robust protocols could still be secure, but completely nonrobust protocols are automaticaly proven insecure \cite{2007Boyer}.

Both BKM2007 and ZQLWL2009 have a common problem: the sources of the photons are attenuated laser pulses which have a nonzero probability to contain two or more photons, leaving such systems subject to the so-called beam splitter attack which has been discussed in Ref. \cite{1992SPN}. Using entangled photon pairs as generated by parametric down-conversion allows us to approximate a conditional single photon source \cite{1995Kwiat} with a high bit rate \cite{2000Jennewein}, and yet a very low probability for producing two pairs simultaneously. And our protocol is more efficient than BKM2007 and as efficient as ZQLWL2009.

In section \Rmnum{2}, we present an efficient SQKD protocol using Bell state which exploiting the features of entangled photon pairs for generating highly secure keys. In section \Rmnum{3}, we prove our protocol being completely robust. Finally, we give a brief discussion and conclusion.

\section{\Rmnum{2}. SCHEME FOR QKD WITH CLASSICAL BOB}
To define our protocol we first introduce the four polarization entangled states $|\phi^{\pm}\rangle = \frac{1}{\sqrt{2}}(|00\rangle \pm |11\rangle)$ and $|\psi^{\pm}\rangle = \frac{1}{\sqrt{2}}(|01\rangle \pm |10\rangle)$, which are created directly using parametric down-conversion by the method described in Ref. \cite{1995Kwiat}. $|\phi^{\pm}\rangle$ and $|\psi^{\pm}\rangle$ are also known as Bell or EPR state. A SQKD protocol using Bell state to construct is described in the following.

$(1)$ Quantum Alice and classical Bob agree on that the Bell state $|\phi^{+}\rangle$ and $|\psi^{+}\rangle$ represent one bit classical information $0$ and $1$, respectively.

$(2)$ Alice prepares an ordered $N = 4n(1+\delta)$ EPR pairs in the state $|\phi^{+}\rangle = \frac{1}{\sqrt{2}}(|00\rangle + |11\rangle)$, where integer $n$ is the desired length of the INFO string and $\delta > 0$ is a fixed parameter. And Alice divides the ordered EPR pairs into two partner-photon sequences $[P_{1}(H), P_{2}(H), P_{3}(H),\cdots, P_{N}(H)]$ and $[P_{1}(T), P_{2}(T), P_{3}(T),\cdots, P_{N}(T)]$. Here $P_{i}(H)$ and $P_{i}(T)$ are the two photons correlated with each other in the $i$th ($i = 1, 2, \cdots, N$) EPR photon pair. We call $[P_{1}(H), P_{2}(H), P_{3}(H),\cdots, P_{N}(H)]$ the home sequence or simply the H-sequence; the another sequence is called travel sequence or the T-sequence for short.

$(3)$ For each qubit in the H-sequence, Alice randomly selects whether to apply the Pauli operation $X = |0\rangle \langle1| + |1\rangle \langle0|$ or do nothing. We notice that by performing the Pauli operation $X$, it transforms the state $|\phi^{+}\rangle$ into $|\psi^{+}\rangle$. Then she stores the H-sequence and sends the T-sequence to Bob through the quantum channel.

$(4)$ For each qubit arriving, Bob chooses randomly either to reflect it (CTRL) or to measure it in the computational basis and resend it in the same state he found (to SIFT it). He records the results of the measurement which is completely secret to any other person other than Bob himself. Qubits are sent one by one, i.e., Alice sends a qubit only after receiving the previous one and Bob resends a qubit immediately after receiving it \cite{2007Boyer}.

$(5)$ Alice uses an $N$-qubit register to save all photons coming back from Bob. Then she tells Bob through a classical channel that she has received the photon sequence.

$(6)$ After hearing from Alice, Bob announces which qubits he chose to CTRL. It is expected that for approximately $\frac{N}{2}$ qbits of T-sequence, Bob chooses randomly to reflect them. We refer to these qbits as $T_{CTRL}$, and the corresponding qubits in the H-sequence is called $H_{CTRL}$. We refer to the qubits Bob chose to SIFT as $T_{SIFT}$, and the correlated partner-photon is $H_{SIFT}$. They abort the protocol if the number of $T_{SIFT}$ bits is less than $2n$; this happens with exponentially small probability.

$(8)$ Alice checks the error rate on the $T_{CTRL}$ in the following way. She makes Bell measurement on the CTRL qubit (qubit in the $T_{CTRL}$) and corresponding home qubit (qubit in the H-sequence) and compares the measurement result and the corresponding initial EPR state, if they are inequal, some errors may happen. If the error rate on the $T_{CTRL}$ is higher than some predefined threshold $P_{CTRL}$, the protocol aborts, otherwise the next step is executed.

$(9)$ Alice measures particles of $H_{SIFT}$ and particles of $T_{SIFT}$ in the computational basis, and chooses ar random $n$ $H_{SIFT}$ to be TEST qbits. She publishes which are the chosen qbits. Bob publishes the measurement results (of $T_{SIFT}$) corresponding to the TEST qubits. Alice compares the results of these measurements. If they are indeed perfectly correlated, Alice and Bob can certain that there is no eavesdropping; otherwise, they abort the protocol.

$(10)$  Alice announces the measurement results of the remaining $H_{SIFT}$. Bob obtains the raw key by comparing the measurement results of $H_{SIFT}$ and $T_{SIFT}$, which is shown clearly in TABLE. \ref{tab:1}. Alice and Bob select the first $n$ raw key to be used as INFO string.
\begingroup
\squeezetable
\begin{table}
\caption{\label{tab:1} Relations of the initial state, measurement results of $H_{SIFT}$ , $T_{SIFT}$ and Raw key.}
\begin{ruledtabular}
\begin{tabular}{lllc}
  Initial state & Measurement result of $H_{SIFT}$ & Measurement result of $T_{SIFT}$ & Raw key \\
  \hline
  $|\phi^{+}\rangle$ & 0 & 0 & 0 \\
  $|\psi^{+}\rangle$ & 0 & 1 & 1 \\
  $|\phi^{+}\rangle$ & 1 & 1 & 0 \\
  $|\psi^{+}\rangle$ & 1 & 0 & 1 \\
\end{tabular}
\end{ruledtabular}
\end{table}
\endgroup

$(11)$ Then Alice announces error correction code (ECC) and privacy amplification (PA); she and Bob use them to extract the $m$-bit final key from the $n$-bit INFO string.
\section{\Rmnum{3}. SECURITY ANALYSIS}
Now, we discuss the security of our protocol. Firstly, when Alice sends $T$-sequence to Bob, an eavesdropper Eve who wants to get the information on the initial states, may intercept this particle and resend a fake particle instead according to her measurement result. Because the state of particle in $T$-sequence is
\begin{eqnarray}
\rho_{T_{i}} = Tr_{H_{i}}(\rho_{H_{i}T_{i}}) = \frac{1}{2}(|0\rangle\langle0| + |1\rangle\langle1|) = \frac{I}{2},
\end{eqnarray}
where $i = 1, 2, 3, \cdots, N$. This state has no dependence upon the initial entangled states, and thus any measurements performs by Eve will contain no information about the initial states, thus preventing Eve from knowing the secret key. And we show that if Eve resends a fake particle to Bob, for example, this fake particle is in the state $|\varphi_{E}\rangle = c|0\rangle + d|1\rangle$, where $|c|^{2} + |d|^{2} = 1$. If Bob chooses to reflect it. The state of the fake particle and the home particle is
\begin{eqnarray}
\rho_{HE} &=& \frac{1}{2}(|0\rangle\langle0| + |1\rangle\langle1|)  \nonumber \\
&&  \otimes (c^{2}|0\rangle\langle0| + cd^{*}|0\rangle\langle1| + c^{*}d|1\rangle\langle0| + d^{2}|1\rangle\langle1|).
\end{eqnarray}
When Alice and Bob make eavesdrop checking, Alice makes a Bell measurement on this fake particle and home particle and she will get any one of four Bell states with equal probability $\frac{1}{4}$. So the error rate introduced by Eve is $\frac{3}{4}$.

Furthermore, the most general attack of Eve can be described by a unitary operator $U_{E}$, which causes one or both of the EPR particles to interact coherently with an auxiliary quantum system available to her for subsequent measurements of her own. The most general global state before Bob decides whether to SIFT or CTRL is of the form
\begin{eqnarray}
|\Phi\rangle &=& |00\rangle_{HT} |A\rangle_{E} + |01\rangle_{HT} |B\rangle_{E} \nonumber \\
&&+ |10\rangle_{HT} |C\rangle_{E} +|11\rangle_{HT} |D\rangle_{E},
\end{eqnarray}
where $|A\rangle$, $|B\rangle$, $|C\rangle$ and $|D\rangle$ are Eve's choices for states of her system, which she does not even have to decide how to measure until after Alice and Bob have gone public.

Suppose the initial state Alice prepares is $|\phi^{+}\rangle$. On the qubit coming back, Eve applies the unitary $U_{E}$; if Bob sifted, the global state before Eve applies $U_{E}$ is $|00\rangle_{HT} |A\rangle_{E} + |11\rangle_{HT} |D\rangle_{E}$. Once Eve has applied $U_{E}$, it must be such that $U_{E}|00\rangle_{HT} |A\rangle_{E} = |00\rangle_{HT} |E_{0}\rangle_{E}$ else the SIFT can detect an error, and similarly $U_{E}|11\rangle_{HT} |D\rangle_{E} = |11\rangle_{HT} |E_{1}\rangle_{E}$. Due to the linearity of quantum mechanics, if Bob reflects (CTRL), the resulting final state must be $U_{E}|\Phi\rangle = |00\rangle_{HT} |E_{0}\rangle_{E} + |11\rangle_{HT} |E_{1}\rangle_{E}$.
As $U_{E}|\Phi\rangle = |\phi^{+}\rangle_{HT}(|E_{0}\rangle_{E} + |E_{1}\rangle_{E}) + |\phi^{-}\rangle_{HT}(|E_{0}\rangle_{E} - |E_{1}\rangle_{E})$ and $|\phi^{-}\rangle_{HT}$ to have probability $0$ of being measured by Alice, $|E_{0}\rangle_{E} = |E_{1}\rangle_{E}$ must hold.

Similarly, if the initial states is $|\psi^{+}\rangle$. On the qubit coming back, Eve applies the unitary $U_{E}$; if Bob sifted, the global state before Eve applies $U_{E}$ is $|01\rangle_{HT} |B\rangle_{E} + |10\rangle_{HT} |C\rangle_{E}$. Once Eve has applied $U_{E}$, it must be such that $U_{E}|01\rangle_{HT} |B\rangle_{E} = |01\rangle_{HT} |E_{1}\rangle_{E}$ else the SIFT can detect an error, and similarly $U_{E}|10\rangle_{HT} |C\rangle_{E} = |10\rangle_{HT} |E_{0}\rangle_{E}$. Due to the linearity of quantum mechanics, if Bob reflects (CTRL), the resulting final state must be $U_{E}|\Phi\rangle = |10\rangle_{HT} |E_{0}\rangle_{E} + |01\rangle_{HT} |E_{1}\rangle_{E}$.
As $U_{E}|\Phi\rangle = |\psi^{+}\rangle_{HT}(|E_{1}\rangle_{E} + |E_{0}\rangle_{E}) + |\psi^{-}\rangle_{HT}(|E_{1}\rangle_{E} - |E_{0}\rangle_{E})$ and $|\psi^{-}\rangle_{HT}$ to have probability $0$ of being measured by Alice, $|E_{0}\rangle_{E} = |E_{1}\rangle_{E}$ must hold.

So the final state is $U_{E}|\Phi\rangle = |\phi^{+}\rangle_{HT}|E_{0}\rangle_{E}$ if the initial state is $|\phi^{+}\rangle_{HT}$ and  $U_{E}|\Phi\rangle = |\psi^{+}\rangle_{HT}|E_{0}\rangle_{E}$ if the initial state is $|\psi^{+}\rangle_{HT}$. Thus the only faked source sure of passing Alice's and Bob's checking is one in which Eve's system is entirely uncorrelated with the EPR particles, so that a subsequent measurement on it tells her no information.

As Eve may gain a certain amount of information without being detected, for example, Eve measures the SIFT bits in computational basis. So to reduce Eve's information to an arbitrarily low value, some privacy amplification protocols are needed. After privacy amplification Alice and Bob would end up with a shared random sequence.
\section{\Rmnum{4}. CONCLUSION}
Our protocol is efficient in that it uses all Bell states in distributing the key except those, approximately half of the Bell states, chosen for checking eavesdropping. This is different from the BKM2007 where only $\frac{3}{4}$ of the particles are used as keys. We now study the efficiency of the protocol. We consider the definition given in Ref. \cite{Cabello2000},
\begin{eqnarray}
\eta = \frac{b_{s}}{q_{t}+b_{t}},
\end{eqnarray}
where $b_{s}$ is the length of the INFO string, $q_{t}$ is the number of transmitted qubits on the quantum channel, and $b_{t}$ is the number of transmitted bits on the classical channel. Here the classical bits used for eavesdrop checking have been neglected.
And the efficiency of our protocol (approximate $\frac{1}{8}$) is higher than that of the BKM2007 protocol (about $\frac{1}{16}$) and not lower than that of the ZQLWL2009 protocol (about $\frac{1}{8}$), which is shown in TABLE. \ref{tab:2}. From a theoretical point of view the scheme provides an interesting and new extension of Boyer et al.'s original idea using Bell state, and give an efficient and secure protocol; but from a practical point of view it may be difficult to realize because building a reliable quantum memory is still a major research goal in experimental quantum physics \cite{2004量子存储,2008量子存储,2010噪音存储模型} and current technology allowing storage time is still limited.
\begingroup
\squeezetable
\begin{table}
\caption{\label{tab:2} Efficiency of BKM2007, ZQLWL2009 and our protocol.}
\begin{ruledtabular}
\begin{tabular}{llllc}
    &$q_{t}$ & $b_{s}$ &$b_{t}$ &Efficiency \\
  \hline
  BKM2007     & $8n$ & $n$ &$8n$&$\frac{1}{16}$ \\
  ZQLWL2009   & $4n$ & $n$ &$4n$ &$\frac{1}{8}$ \\
  Our protocol& $4n$ & $n$ &$4n$&$\frac{1}{8}$ \\
\end{tabular}
\end{ruledtabular}
\end{table}
\endgroup

In conclusion, we present a SQKD protocol using Bell state that is secure against beam splitter attack and more efficient than BKM2007 and as efficient as ZQLWL2009.
\section{\Rmnum{5}. ACKNOWLEDGMENTS}
This work is in part supported by the Key Project of NSFC-Guangdong
Funds (No.U0935002).

\end{document}